\documentclass[reprint,onecolumn,apl,english,10pt,nofootinbib]{revtex4-1}

\usepackage[utf8x]{inputenc}
\usepackage{amsmath}
\usepackage{amssymb}
\usepackage{bbm}
\usepackage[english]{babel}
\usepackage{graphicx}
\usepackage{verbatim}
\usepackage{latexsym}
\usepackage{color}

\newcommand{\dd}[2]{\frac{d #1}{d #2}}
\newcommand{\ddsec}[2]{\frac{d^2 #1}{d #2^2}}
\newcommand{\der}[2]{\frac{\partial #1}{\partial #2}}

\DeclareMathOperator{\arcosh}{arcosh}

\begin{document}

\title{Five dimensional formulation of a DSR}
\author{{\bf Riccardo Junior BUONOCORE}}
\affiliation{Department of Mathematics, King's College London, The Strand, London, WC2R 2LS, UK}
\email{riccardoj.buonocore@gmail.com}

\begin{abstract}
In this paper, we analyze a possible formalization of the deformed special relativity as a five-dimensional theory. This is not the first attempt to do so, but we feel that either these previous treatments are too arbitrary in the choice of the new enlarged space, or they lack a satisfactory physical interpretation. In this work, we propose an algorithm which fixes the shape of the enlarged space. Afterwards, we focus our attention on the consequences of our formalism, proposing a physical interpretation.
\end{abstract}

\maketitle

\newpage

\section{Introduction}
In the last two decades, interest has grown in the introduction of a new observer-independent scale besides the speed of light into a coherent theoretical framework. A turning point has been the demonstration in Ref. \cite{dsr} that the postulates of special relativity can be modified in order to accommodate the presence of the Planck length $\displaystyle\ell=\sqrt{\frac{\hbar G}{c^3}}$. This work has then been followed by Refs. \cite{dsr2} and \cite{dsr3} where generalizations to other constants, such as, respectively, the Plank mass $\displaystyle m=\sqrt{\frac{\hbar c}{G}}$ and the Planck energy $\displaystyle\varepsilon=\sqrt{\frac{\hbar c^5}{G}}$, have been given. Such proposals go by the name of deformed special relativity (DSR) theories. Despite this, the study of a theory where both $\hbar$, the reduced Planck constant, and $G$, the Newtonian universal gravitational constant, are different from zero would clearly require one to solve the whole quantum gravity problem. Recently, Refs. \cite{relativelocality1} and \cite{relativelocality2}, have proposed a theory known as Relative Locality, which tries to create an intermediate step, suggesting the study of a phenomenological limit of the full quantum gravity, defined by the limits $\hbar,G\rightarrow 0$ but holding fixed their ratio. Thus, pure quantum and gravitational effects are turned off, but the possible effects due to
the presence of the Planck mass $m$ are kept. The physical role to give to $m$ remains however still unclear.

Recently, some authors (see Refs. \cite{girelli0,girelli1}) have proposed that a natural way to accommodate the presence of a new observer-independent constant should be to formulate the DSR as a five-dimensional theory. As pointed out by the authors, this formalism would solve technical and interpretative problems such as the nonlinear composition of momenta and the famous soccer ball problem. However, their proposal’s choice and physical interpretation of the fifth dimension have been given in an arbitrary way. Moreover, as underlined in Ref. \cite{rovelli}, the addition of one more dimension imposes the enlargement of the isometry group of the theory with the introduction of new transformations, which apparently lead to a macroscopic relativity in the concept of mass, which is clearly not observed.

Another five-dimensional formulation, by means of a more formal approach, has been given in Refs. \cite{otto0,otto1,otto2,otto3,otto4},where the authors derive a precise shape for the fifth coordinate, despite its unusual dimension. However, in this proposal too, the interpretation of this fifth dimension and the problem of the variation of the mass remain open.

Since we believe that it is necessary to understand first, before going into the study of a full quantum gravity theory, the phenomenological limit proposed in Refs. \cite{relativelocality1,relativelocality2}, and that the introduction of a new observer independent constant should be treated in a five-dimensional setup,\footnote{Just as in the transition from Galilean relativity to Einsteinian relativity, the addition of the observer independent constant $c$ makes time the fourth dimension.}, the aim of this paper is to give a five-dimensional approach to the DSR which should be geometrically well grounded with reasonable phenomenological consequences. In Sec. \ref{intuition1}, we propose an algorithm which will allow us to derive special relativity (with the intuition of the need of the fourth dimension) from Galilean relativity, implementing in the latter the postulate about the role of the speed of light $c$. The reliability of this algorithm is demonstrated in the fact that
we obtain the right theory, i.e., special relativity. In Sec. \ref{intuition2}, we apply this procedure to special relativity itself in order to introduce a second fundamental constant. This constant is the Planck mass $m$, and the role we give to it is to determine the upper bound for the norm of the 4-momentum of the elementary particles. This assumption, despite seeming quite arbitrary, is the same as that found in Ref. \cite{spinning1}, where it arises in the context of 3D gravity. Moreover, since elementary particles are considered to be the building blocks of macroscopic matter, they should not have a rest mass capable of creating a black hole. Section \ref{mass-shell}  is dedicated to the introduction of a mass-shell relation over the new five-dimensional space. In Sec. \ref{the_propagation} we study the shape of the equation of motion of a freely propagating particle, which in turn will allow us to give an interpretation to the fifth dimension. In Secs. \ref{isometrygroup} and \ref{proposal} we analyze systematically the isometry group of our theory, giving a physical interpretation to the transformations. Since the analysis done so far will hold only for elementary particles, in Sec. \ref{soccerball} we discuss the soccer ball problem in our context.

\section{From Galilean Relativity to Einsteinian Relativity}\label{intuition1}
As anticipated in the Introduction, in this section we propose a formal procedure which allows one to deduce the theory of Special Relativity directly by enforcing its postulates into the setup of Galilean relativity, giving particular attention to the request that the speed of light in vacuum must be an upper bound to the speed of the other particles. Let us start, then, with a model of Galilean relativity defined by the couple $(\mathcal E,\delta)$, where $\mathcal E$ is the Euclidean 3-plane equipped with the flat positive-definite metric $\delta$, i.e., $\delta=diag(1,1,1)$. The only postulate holding here is the well-known Galilean relativity principle:
\\
\\
\textit{Postulate 1}: The laws of physics take the same form in every inertial frame.
\\
\\
Let us then introduce a coordinate system $x^i$, which identifies the points of $\mathcal E$ and a parameter $t$ called the universal time. The square modulus of the speed of a
particle is defined as
\begin{equation}\label{Galileian_speed}
v^2=\dd{x^i}{t}\dd{x_i}{t},
\end{equation}
where the Einstein notation is understood and $x^i=x_i$ with $i=1,2,3$. We now introduce the second postulate imposed by special relativity, which can be restated as follows:
\\
\\
\textit{Postulate 2}: The speed of light $c$ in vacuum is the same for every inertial observer, and it is an upper bound for the speed of the other particles.
\\
\\
Thus, for a generic particle with speed $v$, we can write the
condition
\begin{equation}\label{Galileian_constraint}
v^2\leq c^2,
\end{equation}
which can be rewritten as
\begin{equation}\label{Galileian1}
\dd{x^i}{t}\dd{x_i}{t}=c^2\left(1-a^2\right),
\end{equation}
where $a^2\leq 1$ is a function which will depend on the particle we are considering. With some algebra, we can arrange the equation (\ref{Galileian1}) in the following way
\begin{equation}\label{Galileian2}
1-\dd{x^i}{(ct)}\dd{x_i}{(ct)}=a^2
\end{equation}
Now, we note that we can always change the parameter with respect to what we are deriving by means of a diffeomorphism, provided that $\displaystyle \dd{(ct)}{s}>0$, where we call $s$ the new parameter with $[s]=meters$. Let us then multiply both sides of equation (\ref{Galileian2}) by $\displaystyle \left(\dd{(ct)}{s}\right)^2$, obtaining
\begin{equation}\label{Galileian3}
\dd{(ct)}{s}\dd{(ct)}{s}-\dd{x^i}{s}\dd{x_i}{s}=a^2\left(\dd{(ct)}{s}\right)^2.
\end{equation}
Guided by the well-known result, we are led to interpret the quantity $ct$ not as a parameter but as a coordinate in some higher-dimensional space — in this case it is four dimensional — so that the left-hand side of Eq. (\ref{Galileian3}) seems to be the norm of a vector in this four-dimensional space. As an intermediate passage, we then define $x^0\doteq ct$:
\begin{equation}\label{Galileian4}
\dd{x^0}{s}\dd{x^0}{s}-\dd{x^i}{s}\dd{x_i}{s}=a^2\left(\dd{x^0}{s}\right)^2.
\end{equation}
However, the presence of the minus sign between the first and the second terms on the left-hand side does not allow us to consider it as a true norm. With the intuition in mind that it should be some kind of norm and guided again by the well-known result, we could argue that the spatial coordinates ($x^i$ , $i=1, 2, 3$) with subscript indexes bring with them a minus sign, while the temporal coordinate ($x^0$) has the same sign in each case. Thus, defining $x^i=−x^i$ and $x^0=x^0$ we can rewrite Eq. (\ref{Galileian4}) as\footnote{We want to remark that, with this choice, the value of the quantity $\displaystyle \dd{x^i}{t}\dd{x_i}{t}$ changes from $v^2$ to $-v^2$. The other possibility is $x^0=-x_0$ and $x^i=x_i$ which would lead also to a change of sign of the right hand sign of Eq. (\ref{Galileian4}).}
\begin{equation}\label{Galileian5}
\dd{x^\alpha}{s}\dd{x_\alpha}{s}=a^2\left(\dd{x^0}{s}\right)^2,
\end{equation}
where $\alpha$ goes from $0$ to $3$. Since we have shaped the left hand side of equation (\ref{Galileian5}) to be a norm, we are led to infer that the metric of this four dimensional space (now spacetime) is $\eta=diag(1,-1,-1,-1)$. We are then almost able to obtain the special relativistic result; however, we have still to analyze the right-hand side of Eq. (\ref{Galileian5}). It is well known from geometry that, if a curve is parametrized by its curvilinear abscissa, its tangent vectors have unitary norm. If we impose, then, $s$ to be the curvilinear abscissa of the curve $x(s)$ over the spacetime, $a$ must satisfy the condition
 $\displaystyle a=\dd{s}{x^0}$, so that equation (\ref{Galileian5}) can be written as
\begin{equation}\label{Galileian6}
\dd{x^\alpha}{s}\dd{x_\alpha}{s}=1.
\end{equation}
In order to be symmetric with respect to the Galilean case, we want the parameter over the trajectories of the particle to have the dimension of a time, so we introduce a new parameter $\tau$, defining $s=c\tau$ with $[\tau]=seconds$. We underline that in this way we have recovered the usual definition of proper time. Thus we can rewrite Eq. (\ref{Galileian6}) as
\begin{equation}\label{Galileian7}
\dd{x^\alpha}{\tau}\dd{x_\alpha}{\tau}=c^2,
\end{equation}
which is the Special Relativistic constraint over 4-velocities. Summarizing, Eq. (\ref{Galileian_constraint}) imposed over the 3-velocities of physical particles leads us to a theory defined by the couple $(\mathcal M,\eta)$, where $\mathcal M$ is a 4-dimensional spacetime (instead of the 3-space we had before) and $\eta$ is the metric over $\mathcal M$, and the constraint (\ref{Galileian7}) over the 4-velocities of physical particles.
\\

Before closing this section, we find it useful to stress that, once we have found the semi-Riemannian manifold $\mathcal (M,\eta)$ where the motion takes place, we can easily satisfy postulate 1 too. In fact, we could find the isometry group under which the theory is covariant just by computing the Killing vectors, so that we would be able to satisfy the first postulate by formulating laws which are covariant with respect to this group.\footnote{Obviously, after noting that the elements of the isometry group are a generalization of the transformations between inertial observers in Galilean relativity.} However we are not going to do this explicitly, since the result is already well-known. Instead, from the next section onward we are going to use all this machinery in order to try to address the problem of adding one more universal constant to the special relativity.

\section{The emergence of a fifth dimension}\label{intuition2}
In order to apply again the procedure exposed in the previous section, we have to start first by stating the postulate we want to add to the first two:
\\
\\
\textit{Postulate 3}: The Planck mass $m$ is equal for every inertial observer, and the quantity $m^2c^2$ is an upper bound for the squared norm of the 4-momentum of the elementary
particles.
\\
\\
Following, \footnote{Actually, in Ref. \cite{dsr}, the author has been more careful about the second postulate. In fact, paraphrasing him, the speed of light in vacuum should, more generally, be $c$ just in the limit $m\rightarrow \infty$. However we will see that this feature does not apply here.} then, the path of the previous section, we introduce the condition, analogous to Eq. (\ref{Galileian_constraint}),
\begin{equation}\label{condition}
p^\alpha p_\alpha\leq m^2c^2,
\end{equation}
where the momenta are defined by the relation $\displaystyle p_\alpha=m_p\dd{x_\alpha}{\tau}$, with $\tau$ the proper time, $m_p$ the mass of the particle with 4-momentum $p$ and $\alpha=0,1,2,3$. Following Eq. (\ref{Galileian1}), we rewrite Eq. (\ref{condition}) as
\begin{equation}\label{Einstenian1}
m_p^2\dd{x^\alpha}{\tau}\dd{x_\alpha}{\tau}=m^2c^2\left(1-a^2\right),
\end{equation}
where $a^2\leq 1$ is, as before, a function which will depend on the particle under consideration. Let us arrange this equation as
\begin{equation}\label{Einstenian2}
1-\dd{(x^\alpha)}{(\frac{m}{m_p}c\tau)}\dd{(x_\alpha)}{(\frac{m}{m_p}c\tau)}=a^2.
\end{equation}
As we did for the Galilean case, we notice that we can always change the parameter of derivation by means of a diffeomorphism. We will call, then, the new parameter $\Lambda$, with $[\Lambda]=meters$, and we impose as before the condition $\displaystyle\dd{(\frac{m}{m_p}c\tau)}{\Lambda}>0$. Multiplying both sides of Eq.(\ref{Einstenian2}) by $\left(\displaystyle\dd{(\frac{m}{m_p}c\tau)}{\Lambda}\right)^2$, it becomes
\begin{equation}\label{Einstenian3}
\dd{(\frac{m}{m_p}c\tau)}{\Lambda}\dd{(\frac{m}{m_p}c\tau)}{\Lambda}-\dd{(x^\alpha)}{\Lambda}\dd{(x_\alpha)}{\Lambda}=a^2\left(\dd{(\frac{m}{m_p}c\tau)}{\Lambda}\right)^2.
\end{equation}
In the spirit of the manipulations of the previous section, we could interpret the quantity $\displaystyle\frac{m}{m_p}c\tau$ no longer as a parameter but as a coordinate in a higher-dimensional space — five dimensional in this case — so that, again, the left-hand side of Eq. (\ref{Einstenian3}) seems to be some sort of norm. Let us call the enlarged space $\mathcal B$, and let us define then the coordinates in this new space in the following way:
\begin{equation}\label{Einstenian_coordinates}
\chi^\mu=\left\{
\begin{array}{lll}
x^\mu &	&\mu=0,1,2,3\\
\frac{m}{m_p}c\tau	&	&\mu=4,
\end{array}
\right.
\end{equation}
where, clearly, $[\chi^\mu]=meters$. Before going on with our analysis, we notice that in our context the dimension of the fifth coordinate is coherent with the others, so the problem
raised in Refs. \cite{otto0,otto1,otto2,otto3,otto4}, which we have recalled in the Introduction, does not apply here.
\\
In light of Eq. (\ref{Einstenian_coordinates}), we rewrite Eq. (\ref{Einstenian3}) as\footnote{We use convention that the index $\alpha$ runs from $0$ to $3$, while the index $\mu$ runs from $0$ to $4$.}
\begin{equation}\label{Einstenian4}
-\dd{\chi^\alpha}{\Lambda}\dd{\chi_\alpha}{\Lambda}+\dd{\chi^4}{\Lambda}\dd{\chi^4}{\Lambda}=a^2\left(\dd{\chi^4}{\Lambda}\right)^2.
\end{equation}
We have now the same problem we had in Eq. (\ref{Galileian4}): the relative sign between the first and the second terms of the left-hand side of Eq. (\ref{Einstenian4}) does not allow us to consider the whole left-hand side as a norm. Using the same trick used in the previous section, we argue that $\chi^\alpha=-\chi_\alpha$ while $\chi^4=\chi_4$, so\footnote{We note that this condition corresponds to a change of the signature of the old Minkowskian four-dimensional spacetime. This in turn implies that, for example, from now on, the quantity $p^\alpha p_\alpha$ will be negative definite.} Eq. (\ref{Einstenian4}) can now be written as
\begin{equation}\label{Einstenian5}
\dd{\chi^\mu}{\Lambda}\dd{\chi_\mu}{\Lambda}=a^2\left(\dd{\chi^4}{\Lambda}\right)^2.
\end{equation}
Now the left hand side of Eq. (\ref{Einstenian5}) has the shape of a norm in a five-dimensional, flat semi-Riemannian manifold equipped with the metric $g=diag(-1,1,1,1,1)$. If we give again to $\Lambda$ the role of the curvilinear abscissa over the curve $\chi(\Lambda)\subset \mathcal B$, $a$ must satisfy the condition $\displaystyle a=\dd{\Lambda}{\chi^4}$ so that Eq. (\ref{Einstenian5}) becomes
\begin{equation}\label{Einstenian6}
\dd{\chi^\mu}{\Lambda}\dd{\chi_\mu}{\Lambda}=1.
\end{equation}
Finally, we want to describe the evolution of a particle with respect to a parameter which has the dimension of a time, so we define $\displaystyle\Lambda=\frac{m}{m_p}c\lambda$ with $[\lambda]=seconds$. A physical interpretation of $\lambda$ which distinguishes its role from that of the $\tau$ appearing in $\chi^4$ will be given in section \ref{the_propagation}. Calling, then, $\displaystyle\dd{\chi^\mu}{\lambda}$ 5-velocity, its norm computed with respect to $\lambda$ reads as
\begin{equation}\label{Einstenian7}
\dd{\chi^\mu}{\lambda}\dd{\chi_\mu}{\lambda}=\frac{m^2}{m_p^2} c^2,
\end{equation}
which is then a constraint elementary particles have to satisfy in this five-dimensional framework. Before summarizing what we have found in this section, once the 5-velocity is defined, it is straightforward to build the 5-momentum as
\begin{equation}\label{Einstenian_5-momentum}
\Pi_\mu=m_p\dd{\chi_\mu}{\lambda},
\end{equation}
which then satisfies the following dispersion relation:
\begin{equation}\label{Einstenian_dispersion_relation}
\Pi^\mu\Pi_\mu=m^2c^2.
\end{equation}
We are now going to show briefly that the definition (\ref{Einstenian_5-momentum}) leads to the right Special Relativistic limit. In fact, manipulating first the relation in Eq. (\ref{Einstenian7}) we obtain that
\begin{equation}\label{Einstenian_dtau/dlambda}
\dd{\tau}{\lambda}=\frac{1}{\sqrt{1+\frac{p^\alpha p_\alpha}{m^2c^2}}},
\end{equation}
where $p_\alpha$ is defined just like in Special Relativity as $\displaystyle m_p\dd{x_\alpha}{\tau}$. In light of Eq. (\ref{Einstenian_dtau/dlambda}), then, we can rewrite the 5-momentum in the following way:
\begin{equation}\label{Einsteinian-Pi-vs-p}
\begin{array}{lll}
\Pi_\mu
&=&
\displaystyle
\left(m_p\dd{\chi_\alpha}{\lambda},mc\dd{\tau}{\lambda}\right)=

\left(m\dd{\chi_\alpha}{\tau}\dd{\tau}{\lambda},mc\dd{\tau}{\lambda}\right)=

\left(\frac{p_\alpha}{\sqrt{1+\frac{p^\alpha p_\alpha}{m^2c^2}}},\frac{mc}{\sqrt{1+\frac{p^\alpha p_\alpha}{m^2c^2}}}\right)
\end{array}
\end{equation}
It is evident that in the limit $m\rightarrow\infty$, the first four components reduce to the usual 4-momentum\footnote{We notice that the fifth component diverges just like the zero component of the Special Relativistic 4-momentum in the limit $c\rightarrow\infty$.}.
\\

Just as we did at the end of the previous section, we end up with a theory defined by a new couple $(\mathcal B,g)$, where the dimension of the manifold where the motion takes place grows by 1. Relations (\ref{Einstenian7}) and (\ref{Einstenian_dispersion_relation}) are the five-dimensional counterparts of the well-known special relativistic ones and, as shown by Eq. (\ref{Einsteinian-Pi-vs-p}) they define quantities which have the right expected limit. The next step is to find a suitable definition of the mass-shell relation for this theory. We remark that obviously, in what follows, we could have chosen the opposite signature of the metric without changing the results.

\section{The mass-shell relation}\label{mass-shell}
In special relativity, the mass-shell relation arises quite naturally from the definitions of 4-velocity and proper time. However, in $\mathcal B$, there is not such an evident relation between the mass of a particle and its 5-momentum or its 4-momentum. In the context of Relative Locality, see Refs. \cite{relativelocality1} and \cite{relativelocality2}, where the momentum space is supposed to be curved, the value of the mass of a particle with momentum $p$ is defined as the value of the geodesical distance from the origin of the momentum space to the point, over the momentum space itself, identified by the coordinates of $p$. We want now to take this intuition and move it into the context of this paper.
\\

The dispersion relation [Eq. (\ref{Einstenian_dispersion_relation})] is a constraint over the value of the components of the 5-momentum. From a geometrical point of view, it defines a semi-Riemannian submanifold of $T^*\mathcal B$; in particular, it is a four-dimensional hyperboloid, which we will call $\mathcal H$. Then, only a four-dimensional chart is needed in order to parametrize it and, looking at Eq. (\ref{Einsteinian-Pi-vs-p}), it is evident that good candidates as coordinates over $\mathcal H$ are the components of the 4-momentum $p_\alpha$. In light of these observations, we argue that the value of the mass of a particle with 4-momentum $p_\alpha$ is equal to the value of the geodesic distance over $\mathcal H$ between the points $(0,0,0,0,mc)$ (written in the embedding space $T^*\mathcal B$), which we will call origin of $\mathcal H$, and $p$ itself\footnote{Our choice of the point with respect to what we compute the distance appears to be straightforward analyzing Eq. (\ref{Einsteinian-Pi-vs-p}); in fact, it is obtained just posing $p_\alpha=0$.}; explicitly
\begin{equation}\label{Mass-shell_mass_definition}
D^2(0,p)=m_p^2c^2.
\end{equation}
Thus we are giving to $\mathcal H$ partially the same role the curved momentum space has for the Relative Locality\footnote{The parallelism is not complete, since in this context there is no need to introduce any connection which rules the composition of the 4-momenta.}. A straightforward computation of $D(0,p)$ (details can be found in Appendix \ref{appendix1}), used in combination with the definition in Eq. (\ref{Mass-shell_mass_definition}), gives the following relation between the 4-momentum and the mass of a particle:
\begin{equation}\label{Mass-shell_on_shell_relation_section}
p^\alpha p_\alpha=-m^2c^2\tanh^2\left(\frac{m_p}{m}\right),
\end{equation}
which holds for both massive and massless elementary particles. This is then our proposal of mass-shell relation to enforce on $\mathcal B$ which we notice is just a deformation of the
special relativistic one, and in the limit $m\rightarrow\infty$ it reduces to $p^\alpha p_\alpha=-m_p^2c^2$. The careful reader could be worried by the fact that the left hand side of Eq. (\ref{Mass-shell_on_shell_relation_section}) is no longer a scalar in a five-dimensional space; we will deal with this problem in Sec. \ref{isometrygroup}.

\subsection*{A brief comment upon massless particles}
As just stated, the relation in Eq. (\ref{Mass-shell_on_shell_relation_section}) holds for massless particles too. Despite Eq. (\ref{Einstenian_dispersion_relation}) being perfectly well defined, it could seem that Eq. (\ref{Einstenian7}) is not. Actually, this is not the case, since Eq. (\ref{Einstenian7}) is simply stating that massless particles have the
modulus of their 5-velocity infinite. Moreover, by means of the relation in Eq. (\ref{Einstenian_dtau/dlambda}), it can be easily seen that this in turn implies that the modulus of their 4-velocity is zero, in perfect agreement with special relativity.

\section{The propagation of a free particle}\label{the_propagation}
The aim of this section is to give a first phenomenological prediction of the setup developed so far for the propagation of particles; in particular, we will focus our attention on the corrections to the speed of a free elementary particle. In order to do this, we have to derive first the shape of the equation of motion of a freely propagating particle over $\mathcal B$. To be precise, we should first study the group under which the theory should be covariant, i.e.,
which allows us to change between inertial frames. However, we anticipate that it will be, as expected, the isometry group, and thus we will be here satisfied to formulate the law of the propagation of a free particle as a relation involving only 5-vectors.
\\

Let us start observing that the special relativistic equations of motion of a free particle, \textit{i.e.},
\begin{equation}\label{free_eq_of_motionSR}
\ddsec{x^\alpha}{\tau}=0,
\end{equation}
must hold at least in the limit $m\rightarrow\infty$. In light of Eq.(\ref{free_eq_of_motionSR}), we infer that the equations of motion of the first four components of the coordinate system should be
\begin{equation}\label{4-M5eq_of_motion}
\ddsec{\chi^\alpha}{\lambda}=\ddsec{x^\alpha}{\tau}\left(\dd{\tau}{\lambda}\right)^2=0.
\end{equation}
It can be argued that the equations $\displaystyle\ddsec{x^\alpha}{\tau}=0$ could not hold exactly in the context we are analyzing, but in the most general case, on the right hand side, terms of order at least $\mathcal O(m^{-1})$ could appear. However, in the transition from Galilean relativity to special relativity, the equations $\displaystyle\ddsec{x^i}{t}=0$ keep holding exactly, so arguing that the same feature holds here, we impose that $\displaystyle\ddsec{x^\alpha}{\tau}=0$ on $\mathcal B$ too.
\\
For the fifth coordinate, we notice that the constraint in Eq. (\ref{Einstenian7}) must always hold, so, combining it with Eq. (\ref{4-M5eq_of_motion}), we deduce that $\displaystyle\dd{\chi^4}{\lambda}$ must also be a constant during a free motion, \textit{i.e.},
\begin{equation}
\ddsec{\chi^4}{\lambda}=0.
\end{equation}
Summarizing, the equations of motion of a freely propagating particle over $\mathcal B$ are simply
\begin{equation}\label{M5eq_of_motion_chi}
\ddsec{\chi^\mu}{\lambda}=0.
\end{equation}
It is easy to deduce from Eq. (\ref{M5eq_of_motion_chi}) the conservation of the 5-momentum. In fact, taking into account the definition (\ref{Einstenian_5-momentum}), Eq. (\ref{M5eq_of_motion_chi}) can be rewritten as
\begin{equation}\label{M5eq_of_motion_Pi}
\dd{\Pi_\mu}{\lambda}=0.
\end{equation}
Some kind of link between Eq. (\ref{M5eq_of_motion_Pi}) and the well-known conservation of the 4-momentum of the special relativity is then expected. We have already established that the first four components of the 5-momentum reduce to the special relativistic 4-momentum in the limit $m\rightarrow\infty$ (cfr. Eq. (\ref{Einsteinian-Pi-vs-p})). It can be easily inferred, then, that the first four components of Eq. (\ref{M5eq_of_motion_Pi}) are simply stating a generalization of the conservation of the 4-momentum\footnote{In special relativity, in the  limit $c\rightarrow\infty$, the spatial part of the 4-momentum reduces to the Galilean 3-momentum in the same way.}. For what concerns the fifth component of Eq. (\ref{M5eq_of_motion_Pi}), it is instead stating a new conservation law, explicitly
\begin{equation}\label{M5new_conservation}
\dd{\Pi_4}{\lambda}=\dd{}{\lambda}\left(\frac{mc}{\sqrt{1+\frac{p^\alpha p_\alpha}{m^2c^2}}}\right)=0.
\end{equation}
Since the only dynamical quantities between the parentheses are the $p_\alpha$'s, taking into account the relation in Eq. (\ref{Mass-shell_on_shell_relation_section}), it follows that the equation of motion [Eq. (\ref{M5new_conservation})] is stating that in a free motion, the mass of an elementary particle does not change\footnote{Despite the fact that empirical results on the neutrino oscillation seem to go against this, we remark that this theory is still classical, since $\hbar\rightarrow 0$; thus, purely quantum effects are not taken into account.}.
\\

Now we have all the instruments to focus our attention on the speed of a freely propagating particle, say, along the positive $\chi^1$ direction with spatial speed $v$; then $\Pi_2=\Pi_3=p_2=p_3=0$. The following chain of equalities holds:
\begin{equation}\label{M5_speed1}
\frac{v}{c}=\frac{\dot\chi^1}{\dot\chi^0}=\frac{\Pi^1}{\Pi^0}=-\frac{\Pi_1}{\Pi_0}=-\frac{p_1}{p_0},
\end{equation}
where the dot means the derivation with respect to $\lambda$, and the second equality follows from the definition in Eq. (\ref{Einstenian_5-momentum}). Now we use the mass-shell relation [Eq. (\ref{Mass-shell_on_shell_relation_section})] derived in the previous section, which can be rewritten, making explicit the contraction over $\alpha$, as
\begin{equation}\label{M5_p0}
p_0=-\sqrt{(p_1)^2+m^2c^2\tanh^2\left(\frac{m_p}{m}\right)},
\end{equation}
where the presence of the minus sign depends on the signature of the metric. Substituting into Eq. (\ref{M5_speed1}) the value of $p_0$ found in Eq. (\ref{M5_p0}), we obtain that
\begin{equation}\label{M5_speed2}
\frac{v}{c}=\frac{p_1}{\sqrt{(p_1)^2+m^2c^2\tanh^2\left(\frac{m_p}{m}\right)}}.
\end{equation}
Finally, in order to confront more easily this value with the special relativistic one, we develop the denominator in powers of $m^{-1}$ up to the first non zero correction, obtaining
\begin{equation}\label{M5speed2-leadingorder}
\frac{v}{c}=\frac{p_1}{\sqrt{(p_1)^2+m_p^2c^2}}\left[1+\frac{1}{3}\frac{m_p^2}{m^2}\frac{m_p^2c^2}{(p_1)^2+m_p^2c^2}\right].
\end{equation}
Evidently, the second term in the square parentheses is a correction to the special relativistic result, which is unfortunately extremely small to be measured. However,
from a theoretical point of view, this result protects the setup formulated so far from being a simple nonlinear reformulation of special relativity. Finally, we notice that, since the mass-shell relation [Eq. (\ref{Mass-shell_on_shell_relation_section})] holds for both massive and massless particles, the speed of the latter can be computed just by posing $m_p=0$ in Eq. (\ref{M5_speed2}). In this case, it reduces exactly to the special relativistic result; \textit{i.e.}, massless particles travels at $c$. This in turn implies that the context we propose here does not predict any kind of delay, with respect to special relativity, in the detection of massless particles.

\subsection*{On the difference between $\tau$ and $\lambda$}
Now that we have both the mass-shell relation and the equation of propagation of a free particle, we are ready to give an interpretation to the fifth dimension. We recall that its definition is
\begin{equation}
\chi^4=\frac{m}{m_p}c\tau.
\end{equation}
In special relativity, $c\tau$ is the length of the path traveled by the particle on the spacetime computed from a certain point fixed by the observer. The free equation of motion for $\chi^4$, then, is
\begin{equation}
\ddsec{\chi^4}{\lambda}=\frac{mc}{m_p}\ddsec{\tau}{\lambda}=0.
\end{equation}
Using Eq. (\ref{Einstenian_dtau/dlambda}) and the mass-shell relation [Eq. (\ref{Mass-shell_on_shell_relation_section})], it follows that
\begin{equation}\label{tauvslambda1}
\dd{\tau}{\lambda}=\frac{1}{\sqrt{1-\tanh^2\left(\frac{m_p}{m}\right)}}=const,
\end{equation}
which means that $\tau$ flows faster then $\lambda$ by a factor which depends on the mass of the particle we are considering. We now want to take the trace $\delta$ of a particle over $\mathcal B$ and compute the length of its projection over a slice of $\mathcal B$, namely a spacetime. Thus, we consider the following quantity:
\begin{equation}
L[\delta|_\mathcal{M}]=
\int_a^bd\lambda\sqrt{\left\vert\dd{x^\alpha}{\lambda}\dd{x_\alpha}{\lambda}\right\vert}=
\int_a^bd\lambda\sqrt{\left\vert\dd{x^\alpha}{\tau}\dd{x_\alpha}{\tau}\right\vert}\left\vert\dd{\tau}{\lambda}\right\vert=
\frac{m}{m_p}c\tanh\left(\frac{m_p}{m}\right)\frac{\Delta\lambda}{\sqrt{1-\tanh^2\left(\frac{m_p}{m}\right)}},
\end{equation}
where in the third equality we have used Eqs. (\ref{Mass-shell_on_shell_relation_section}) and the (\ref{tauvslambda1}). Finally, again using Eq. (\ref{tauvslambda1}), we can write
\begin{equation}\label{Mlenght}
L[\delta|_\mathcal{M}]=\frac{m}{m_p}\tanh\left(\frac{m_p}{m}\right)c\Delta\tau.
\end{equation}
Since in the limit $m\rightarrow\infty$ this expression reduces to $L[\delta|_\mathcal{M}]=c\Delta\tau$, we can infer that the $c\tau$ appearing in $\chi^4$ is the spacetime distance traveled by the observed particle when one neglects the effects due to the mass of the particle itself. It seems natural at this point to give to $\lambda$ the role of a
true proper time, meaning that it is the time actually “measured” by the observed particle. A similar result can be found in an observation done in Ref. \cite{rovelli}, where the author, looking for a purely DSR effect, combines the Compton length of a particle with the fact that, according to general relativity, the mass of a particle slows its proper time. We finish this section by noting that this feature, found using quantum and gravitational arguments, arises
naturally in our formalism.

\section{The isometry group of $\mathcal B$}\label{isometrygroup}
The first postulate of the theory we are here proposing states that the laws of physics must be the same in all inertial reference systems. In special relativity, one deduces that the mutually inertial systems are the ones related by the isometry transformations of the Minkowski spacetime. This deduction is achieved because these transformations are exactly, or a deformation of, the transformations which link mutually inertial frames in classical mechanics. One of the goals of this section is then to demonstrate that the isometry
group of $(\mathcal B,g)$ plays the same role as the Poincaré group in special relativity.
\\

As observed at the end of Sec. \ref{intuition1}, once a manifold with its metric is specified, the analysis of its symmetries is straightforward using the Killing vectors which, we recall briefly, are defined by the relation
\begin{equation}
L_\xi g=0,
\end{equation}
where $L$ is the Lie derivative and $\displaystyle\xi=\xi^\mu(\chi)\der{}{\chi^\mu}$ is a Killing vector field, in this case over $\mathcal B$. In coordinates, this equation reads as
\begin{equation}\label{Killing_equation}
\xi_{\mu;\nu}+\xi_{\nu;\mu}=0,
\end{equation}
where the semicolon stands for the covariant derivative. Despite the fact that the fifth coordinate $\chi^5$ is not strictly a spatial coordinate, geometrically the metric $g$ is nonetheless flat, so the Killing equation [Eq. (\ref{Killing_equation})] reduces to
\begin{equation}\label{flat_Killing_equation}
\xi_{\mu,\nu}+\xi_{\nu,\mu}=0,
\end{equation}
where the comma stands for the ordinary derivative. The solution of Eq. (\ref{flat_Killing_equation}) is then the well-known expression
\begin{equation}	\label{flat_Killing_vectors}
\xi^\mu(\chi)=\Lambda^\mu_{\phantom\mu \nu}\chi^\nu+d^\mu,
\end{equation}
where $\Lambda_{\mu\nu}$ is an antisymmetric matrix and $d^\mu$ is a constant vector with $[\Lambda _{\mu\nu}]=1$ and $[d^\mu]=meters$. A first observation is that, since $b^\mu$ is a vector in a five-dimensional space and $\Lambda$ is a $5 \times 5$ antisymmetric matrix, there will be $15$ independent transformations instead of the ten of special relativity. In order to classify these transformations, we use the fact that the Killing vectors can be thought of as
infinitesimal transformations, so that we can write
\begin{equation}\label{infinitesimal_transformation}
\chi'^\mu=\chi^\mu+\sigma\xi^\mu,
\end{equation}
where $\sigma$ is the parameter of the transformation and $\chi'^\mu$ is the value of the coordinate after the transformation. We will then solve the integral curve equation
\begin{equation}\label{Killing_integral_curves}
\dd{\chi'^\mu(\sigma)}{\sigma}=\xi^\mu
\end{equation}
to obtain the finite transformations. Before we begin with their systematic analysis, we give a parametrization of the $\Lambda$ matrix which will turn out to be very useful in the classification, explicitly
\begin{equation}\label{infinitesimal_transformation_rewritten}
\chi'^\mu=\left[g^{\mu\nu}+\sigma\left(b^\mu a^\nu-a^\mu b^\nu\right)\right]\chi_\nu,
\end{equation}
where $a^\mu$ and $b^\mu$ are both independent five-dimensional vectors of $\mathcal B$. Thus, the interpretation will rely on the proper choice of such vectors.

\subsection{Translations}
The constant vector $d^\mu$ clearly induces a translation over $\mathcal B$. There is no need, then, to pass through the infinitesimal transformation, and we can just write down the finite result as
\begin{equation}\label{generalized_translations}
\chi'^\mu=\chi^\mu+d^\mu.
\end{equation}
The first four components of $d^\mu$ clearly produce the usual spacetime translations, while for the fifth a little bit more attention is needed. For a generic $d^4$, the fifth component of Eq. (\ref{generalized_translations}) reads as
\begin{equation}\label{generalized_translation_d4}
\chi'^4=\chi^4+d^4.
\end{equation}
Since $\displaystyle\chi^4=\frac{m}{m_p}c\tau$, at first sight it could seem that this fifth translation may induce some kind of change to the mass of the particle we are following. We argue instead that the interpretation is the simplest possible: according to us, $d^4$ is just a translation of the proper time, \textit{i.e.}, a change of the point on the worldline of the particle from which we are computing the proper time. Explicitly, we assume that for a
particle with mass $m_p$, the parameter $d^4$ should be interpreted as
\begin{equation}
d^4=\frac{m}{m_p}c\Delta,
\end{equation}
where $\Delta$ is the translation factor over the proper time.

\subsection{Rotations}
Let us now analyze the transformations induced by the matrix $\Lambda_{\mu\nu}$. As said before, we just have to focus on the transformations induced by different choices of the vectors $a^\mu$ and $b^\mu$. We start with the choices $a^\mu=(0,1,0,0,0)$ and $b^\mu=(0,0,1,0,0)$. The infinitesimal transformation [Eq. (\ref{infinitesimal_transformation_rewritten})] reads explicitly as
\begin{equation}\label{chi1chi2rotation}
\left\{
\begin{array}{l}
\chi'^0=\chi^0\\
\chi'^1=\chi^1-\sigma \chi^2\\
\chi'^2=\chi^2+\sigma \chi^1\\
\chi'^3=\chi^3\\
\chi'^4=\chi^4.
\end{array}
\right.
\end{equation}
If we integrate it, it is easy to see that the transformation in Eq. (\ref{chi1chi2rotation}) is a rotation around the $\chi^3$ axis. Thus, giving to $\sigma$ the role of an angle and renaming $\sigma=\vartheta$, the finite transformation is
\begin{equation}\label{M5rotation}
\left\{
\begin{array}{l}
\chi'^0=\chi^0\\
\chi'^1=\chi^1\cos\vartheta-\chi^2\sin\vartheta\\
\chi'^2=\chi^2\cos\vartheta+\chi^1\sin\vartheta\\
\chi'^3=\chi^3\\
\chi'^4=\chi^4.
\end{array}
\right.
\end{equation}
The choices  $a^\mu=(0,1,0,0,0)$, $b^\mu=(0,0,0,1,0)$ and $a^\mu=(0,0,1,0,0)$, $b^\mu=(0,0,0,1,0)$, will obviously lead, respectively, to the spatial rotations around the $\chi^2$ and $\chi^1$ axes.

\subsection{Boosts}
Our next choice is $a^\mu=(1,0,0,0,0)$ and $b^\mu=(0,1,0,0,0)$, so the infinitesimal transformation [Eq. (\ref{infinitesimal_transformation_rewritten})] becomes 
\begin{equation}\label{chi1boost}
\left\{
\begin{array}{l}
\chi'^0=\chi^0-\sigma \chi^1\\
\chi'^1=\chi^1-\sigma \chi^0\\
\chi'^2=\chi^2\\
\chi'^3=\chi^3\\
\chi'^4=\chi^4.
\end{array}
\right.
\end{equation}
The integration of Eq. (\ref{chi1boost}) allows us to interpret it, at least formally, as a boost along the $\chi^1$ axis. Letting $\sigma$ be the rapidity and renaming $\sigma=\psi$, the finite transformation reads as
\begin{equation}\label{M5boosthyperbolic}
\left\{
\begin{array}{l}
\chi'^0=\chi^0\cosh\psi-\chi^1\sinh\psi\\
\chi'^1=\chi^1\cosh\psi-\chi^0\sinh\psi\\
\chi'^2=\chi^2\\
\chi'^3=\chi^3\\
\chi'^4=\chi^4.
\end{array}
\right.
\end{equation}
Finally, rewriting it in terms of the parameters $\displaystyle \gamma\beta=\sinh\psi$ and $\displaystyle \gamma=\cosh\psi$, with $\displaystyle\gamma=\frac{1}{\sqrt{1-\beta^2}}$, it becomes
\begin{equation}\label{M5boost}
\left\{
\begin{array}{l}
\chi'^0=\gamma\left(\chi^0-\beta\chi^1\right)\\
\chi'^1=\gamma\left(\chi^1-\beta\chi^0\right)\\
\chi'^2=\chi^2\\
\chi'^3=\chi^3\\
\chi'^4=\chi^4.
\end{array}
\right.
\end{equation}
Since we are not able to find good reason to give it a different physical meaning with respect to the one it has in special relativity, we infer that $\displaystyle\beta=\frac{v}{c}$. The choices $a^\mu=(1,0,0,0,0)$, $b^\mu=(0,0,1,0,0)$ and $a^\mu=(1,0,0,0,0)$, $b^\mu=(0,0,0,1,0)$ lead us to the boosts in the $\chi^2$ and $\chi^3$ directions, respectively.
\\
Since we have so far recovered translations, rotations, and boosts, we can argue that the isometry group of $(\mathcal B,g)$ encloses the transformations which link mutually inertial frames. Before moving to the new set of transformations, it is worth noting that the mass-shell relation [Eq. (\ref{Mass-shell_on_shell_relation_section})] is covariant under the action of the transformations found so far.

\subsection{Momentum boosts}
This subsection is dedicated to the last four transformations, which in literature (see Refs. \cite{otto0,otto1,otto2,otto3,otto4}) have been referred to as momentum boosts. As we will see, from a formal point of view they are not strictly boosts, but we will keep this name. In this subsection, we are only going to give their formal definition, leaving our physical interpretation of them to the next section. Let us then choose $a^\mu=(1,0,0,0,0)$, $b^\mu=(0,0,0,0,1)$. The infinitesimal transformation [Eq. (\ref{infinitesimal_transformation_rewritten})] then becomes
\begin{equation}\label{M5boost2infinitesimal}
\left\{
\begin{array}{l}
\chi'^0=\chi^0-\sigma \chi^4\\
\chi'^1=\chi^1\\
\chi'^2=\chi^2\\
\chi'^3=\chi^3\\
\chi'^4=\chi^4-\sigma \chi^0.
\end{array}
\right.
\end{equation}
Renaming $\sigma=\psi'$ and integrating, the finite transformation reads as
\begin{equation}\label{M5boost2hyperbolic}
\left\{
\begin{array}{l}
\chi'^0=\chi^0\cosh\psi'-\chi^4\sinh\psi'\\
\chi'^1=\chi^1\\
\chi'^2=\chi^2\\
\chi'^3=\chi^3\\
\chi'^4=\chi^4\cosh\psi'-\chi^0\sinh\psi'.
\end{array}
\right.
\end{equation}

The next choice is $a^\mu=(0,1,0,0,0)$, $b^\mu=(0,0,0,0,1)$, which induces the infinitesimal transformation
\begin{equation}\label{M5rotation2infinitesimal}
\left\{
\begin{array}{l}
\chi'^0=\chi^0\\
\chi'^1=\chi^1-\sigma \chi^4\\
\chi'^2=\chi^2\\
\chi'^3=\chi^3\\
\chi'^4=\chi^4+\sigma \chi^1.
\end{array}
\right.
\end{equation}
Its finite version, renaming $\sigma=\vartheta'$, reads as
\begin{equation}\label{M5rotation2}
\left\{
\begin{array}{l}
\chi'^0=\chi^0\\
\chi'^1=\chi^1\cos\vartheta'-\chi^4\sin\vartheta'\\
\chi'^2=\chi^2\\
\chi'^3=\chi^3\\
\chi'^4=\chi^4\cos\vartheta'+\chi^1\sin\vartheta'.
\end{array}
\right.
\end{equation}
We notice that this transformation formally is a rotation. The choices $a^\mu=(0,0,1,0,0)$, $b^\mu=(0,0,0,0,1)$ and $a^\mu=(0,0,0,1,0)$, $b^\mu=(0,0,0,0,1)$ give us the other two possible rotations, involving in place of the $\chi^1$ axis, the $\chi^2$ and $\chi^3$ axes, respectively.

\section{Proposal of interpretation of the momentum boosts}\label{proposal}
The physical interpretation of the Lorentz boost can be easily achieved by analyzing its infinitesimal version, because its spatial part reduces to the well-known Galilean boost. Thus, one infers that the Lorentz boost codifies the deformations induced by the change between systems of reference with a relative speed. In order to propose a physical interpretation of the momentum boost, we then focus our attention on the transformations in
Eqs. (\ref{M5boost2infinitesimal}) and (\ref{M5rotation2infinitesimal}), which we report here in a synthetic way:
\begin{equation}\label{M5boost2synthetic0}
\left\{
\begin{array}{l}
\chi'^0=\chi^0-\psi' \chi^4\\
\chi'^4=\chi^4-\psi' \chi^0,
\end{array}
\right.
\end{equation}
\begin{equation}\label{M5boost2synthetic1}
\left\{
\begin{array}{l}
\chi'^1=\chi^1-\vartheta' \chi^4\\
\chi'^4=\chi^4+\vartheta' \chi^1.
\end{array}
\right.
\end{equation}
Here\footnote{Needless to say, that the other two possibilities behave exactly as Eq. (\ref{M5boost2synthetic1}).} the task of the identification of a physical interpretation is more complicated, because we do not know yet what is the role of $\psi'$ and $\vartheta'$. What we know is that they must be the ratio of a physical quantity, which defines the property that distinguishes between the new and the old systems of reference, and a relevant scale of the theory. An obvious choice for the scale is clearly the quantity $mc$, which is the one we have introduced in order to deform the special relativity in Sec. \ref{intuition2}. Moreover, since the new transformations are exactly four, it seems reasonable to choose the physical quantities at the numerators to be the components of a 4-vector with the dimension of a 4-momentum. At this point, it seems evident that these transformations will deal with the change of the 4-momentum of the reference frame. In order to clarify this, we are now going to discuss an explicit example, in a special relativistic context, where the change of momentum of the reference frame is implied, creating a link with Eqs. (\ref{M5boost2synthetic0}) and (\ref{M5boost2synthetic1}). We will then discuss the generalizations.

\subsection{A simple example}
Let us consider in a special relativistic context a scattering between two elementary particles with masses $m_p$ and $m_q$, 4-velocities before the collision $u$ and $w$ and
after $u'$ and $w'$ , respectively. Let us underline that the two particles before and after the collision are in a free motion. Since we are dealing with a scattering, the masses of the
two particles do not change, and the process is characterized by the conservation law
\begin{equation}\label{M5process}
p+q=p'+q',
\end{equation}
where $p=m_pu$, $q=m_qw$, $p'=m_pu'$ and $q'=m_qw'$. We now rearrange Eq. (\ref{M5process}), making explicit its dependence upon masses and 4-velocities, as
\begin{equation}\label{M5process'}
u'=u-\frac{m_q}{m_p}(w'-w).
\end{equation}
It can be shown (see Appendix \ref{appendix2}) that this relation can be manipulated into
\begin{equation}\label{M5process_manipulated}
x'=x-\frac{m_q}{m_p}(w'-w)\tau,
\end{equation}
where $x'=u'\tau$, $x=u\tau$ and $\tau$ is the proper time of the particle with mass $m_p$.  As derived in the Appendix \ref{appendix2}, the interpretation of this formula is straightforward: the position of the particle with mass $m_p$ after the collision is the one it would have if the collision had not occurred, plus a correction given by the second member on the right-hand side. Since we are in a special relativistic context, Eq. (\ref{M5process_manipulated}), which has been derived using 4-vectors, holds in any reference frame; in particular, it holds in the reference frame attached to the particle with mass $m_q$.  Giving, then, to the particle with mass $m_q$ the role of a massive reference frame, it is evident that the second member on the right-hand side of Eq. (\ref{M5process_manipulated}) encloses exactly the change of momentum of the reference frame we were looking for. If we then assume, without loss of generality, that after the collision the massive reference frame travels along the $x^1$ axis (in its coordinatization before the collision), defining
\begin{align}
&\psi'=\frac{m_q(w'^0-w^0)}{mc},\\
&\vartheta'=\frac{m_q(w'^1-w^1)}{mc}
\end{align}
and using the definition in Eq. (\ref{Einstenian_coordinates}), the transformations [Eqs. (\ref{M5boost2synthetic0}) and (\ref{M5boost2synthetic1})] read as
\begin{equation}\label{M5boost2synthetic0'}
\left\{
\begin{array}{l}
\displaystyle
x^0=x^0-\frac{m_q}{m_p}(w'^0-w^0)\tau\\
\displaystyle
\tau'=\tau-\frac{m_pm_q(w'^0-w^0)}{m^2c^2}x^0,
\end{array}
\right.
\end{equation}
\begin{equation}\label{M5boost2synthetic1'}
\left\{
\begin{array}{l}
\displaystyle
x^1=x^1-\frac{m_q}{m_p}(w'^1-w^1)\tau\\
\displaystyle
\tau'=\tau+\frac{m_pm_q(w'^1-w^1)}{m^2c^2}x^1,
\end{array}
\right.
\end{equation}
which in the limit $m\rightarrow\infty$ reduce exactly to Eq. (\ref{M5process_manipulated}). According to this result, we are led to interpret the momentum boost as the transformation which codifies the deformations in the coordinatization of a reference frame due to the change of the 4-momentum of the reference frame itself.

\subsection{Further considerations}
The interpretation we gave at the end of the previous subsection gives rise to many questions. The first is that the transformations in Eqs. (\ref{M5boost2synthetic0'}) and (\ref{M5boost2synthetic1'}),  as we have derived them, hold in every (Lorentz) boosted system of reference, thus causing the values of $\psi'$ and $\vartheta'$, which define the strength of the deformations, to be not uniquely defined. We can easily circumvent this problem by stating that their value is the one measured in a system attached to the massive reference frame before the collision, and which afterwards keeps on going in the same direction. In order to clarify this, we can give an explicit formula: the 4-momentum of the massive reference frame before the collision, measured by a system attached to it, is clearly $q=(m_qc,0)$. After the collision, it will be $q'=(m_q\gamma c, m_q\gamma v)$, where\footnote{This is true even in our five-dimensional formulation, since the Lorentz boost is undeformed.} $v$ is the spatial speed it will acquire and $\gamma$ is the usual Lorentz factor; thus
\begin{align}
&\psi'=\frac{m_q(\gamma-1)c}{mc},\\
&\vartheta'=\frac{m_q\gamma v}{mc}.
\end{align}
A second issue is that Eqs. (\ref{M5boost2synthetic0'}) and (\ref{M5boost2synthetic1'}) give two different transformations of the isometry group, but they both have to be applied in order to give the right special relativistic limit. Because of this, it could seem that these two transformations are no longer independent. Actually, this is not the case. In fact, the transformations of the momentum boost are still independent. Moreover, taken separately, they can cause macroscopic changes in the masses of the particles, as can be easily seen by analyzing Eq. (\ref{Mass-shell_on_shell_relation_section}), and as it was already pointed out in Ref. \cite{rovelli}. This problem can be solved by noting that the changes in momentum of massive reference frames are not arbitrary, because they follow the conservation laws. Thus, when applied to physical systems, the transformations of the momentum boost are forced by the conservation laws not to be taken into account separately, so no macroscopic change of the masses is implied. Furthermore, we notice that in the derivation performed in the previous subsection, all the particles are on shell by construction. We have to underline, however, that the transformations in Eqs. (\ref{M5rotation2}) and (\ref{M5boost2hyperbolic}) do not commute but, at this stage, the question about which one should be applied first, can be only answered by experiments.

A third problem is that after the collision, the massive reference frame has acquired a spatial speed; thus, we cannot say we are still attached to it, and thus we are no longer on a massive reference frame. We then postulate that a momentum boost should be followed by the usual Lorentz boost in the direction and with the speed that the massive reference frame has acquired after the collision. We find this consideration particularly interesting, since it
allows the momentum boost to be considered as a deformation of the Lorentz boost.

Another concern arises about what happens to the coordinatization of the particles which are not involved in the collision. The answer is again found in the way we have derived the special relativistic limit of the momentum boost. Since a particle which is not involved in the collision does not change its momentum, its contribution in Eq. (\ref{M5process}) simply cancels out. Thus, the deformations due to the momentum boost occur only for the coordinatization of a particle which exchanges its momentum with the particle reference frame.

Before concluding this subsection, we note that it could seem that in some way we have enlarged the class of the property which defines two mutually inertial frames. Actually, this is not the case: the Galilean definition of mutually inertial frames\footnote{We recall the definition: Two frames are mutually inertial if they are standing still or traveling at a constant speed with respect to each other.} keeps holding, since the massive reference system does change its speed after the collision. The momentum boost simply adds a further specification of the change of its state of motion.

\subsection{Generalizations}
So far, we have analyzed the application of the momentum boost to a very specific case: the scattering. In this subsection, we want to discuss the possible generalizations of its applications, recalling that we are always considering the case of elementary particles. A first generalization could be that the two particles before the collision are different from the particles after because, for example, in a quantum scenario, an exchange of quantum numbers between the particles has occurred. Even though this could be allowed in a classical setup, we notice that in a quantum perspective the massive reference frame loses its identity after the collision, so it is no longer identifiable. In fact, the choice of the new massive reference frame after the collision would be totally arbitrary. Thus, we feel we can exclude such processes from the possible extension of applicability. In fact, despite $\hbar\rightarrow 0$, the strength of the momentum boost is ruled by $m$, which should enclose to some extent quantum features. We notice that this restriction in turn constrains the other particle (in the binary collision) to keep its nature too.

Finally, we make a brief comment on collisions of more than two particles. If we consider a collision, for example, of three particles, the conservation law would be
\begin{equation}
p+q+k=p'+q'+k'.
\end{equation}
Choosing the particle with momentum $p$ to be our massive reference frame, if one performs the same manipulations shown in Appendix \ref{appendix2}, it can be easily seen that this process would lead to different transformations for the particles with momenta q and k. Namely, the deformation parameters for the particles with momenta $q$ and $k$ will be, respectively,
\begin{equation}
\begin{array}{cc}
\displaystyle\sigma_q=\frac{\Delta k+\Delta p}{mc},	&	\displaystyle\sigma_k=\frac{\Delta q+\Delta p}{mc},
\end{array}
\end{equation}
where we have chosen not to specify the components of the momenta. This is not acceptable, since, for example, a law involving the 5-momenta of different particles would not be covariant, because every 5-momentum would change with a different law. Thus, a multiparticle collision cannot be treated by the momentum boost. It is worth noting, however, that collisions of more than two particles are extremely rare.

\section{The soccer-ball problem}\label{soccerball}
The soccer ball problem is an issue which, as has been pointed out, arises in theories that attempt to add another observer-invariant scale, and it has happened in DSR theories (see, for example, Refs. \cite{soccerballMaggiore}, \cite{soccerballHossenfelder}). Briefly, the deformations introduced by the second observer-independent scale are usually related to the Planck energy or (as it is in our case) to the Planck mass. Since macroscopic bodies have energies and masses far beyond these scales, the effects of the deformations should be easily seen for composite objects. As has already been underlined, in Refs. \cite{girelli0} and \cite{girelli1}, the authors pointed out that this problem can be solved in their five-dimensional approach to the DSR, since the deformation scale is not the same for
elementary particles and composite bodies. Nonetheless, a solution has recently been given in the Relative Locality approach too; see, for example, Ref. \cite{soccerballAmelino}. In this section, we make a simple explicit computation showing that in our formulation, the soccer ball problem can still be solved.

Let us then consider the relation in Eq. (\ref{Einstenian_dispersion_relation}), which we report here for clarity:
\begin{equation}\label{soccerball-Dispersion}
\Pi^\mu\Pi_\mu=m^2c^2.
\end{equation}
As we have derived it, this relation should hold only for elementary particles. Let us then consider a body made of two of them with 5-momenta $\Pi$ and $\Gamma$ which both satisfy Eq. (\ref{soccerball-Dispersion}). We parametrize them, following Eq. (\ref{Einsteinian-Pi-vs-p}), as
\begin{align}
&\Pi_\mu=\left(\frac{p_\alpha}{\sqrt{1+\frac{p^\alpha p_\alpha}{m^2c^2}}},\frac{mc}{\sqrt{1+\frac{p^\alpha p_\alpha}{m^2c^2}}}\right)=(\gamma'_pp_\alpha,\gamma'_pmc),\\
&\Gamma_\mu=\left(\frac{q_\alpha}{\sqrt{1+\frac{q^\alpha q_\alpha}{m^2c^2}}},\frac{mc}{\sqrt{1+\frac{q^\alpha q_\alpha}{m^2c^2}}}\right)=(\gamma'_qq_\alpha,\gamma'_qmc),
\end{align}
where we have called $\displaystyle\gamma'_p=\frac{1}{\sqrt{1+\frac{p^\alpha p_\alpha}{m^2c^2}}}$. The total 5-momentum of the composite particle will then be $\Sigma_\mu=\Pi_\mu+\Gamma_\mu$, whose norm is
\begin{equation}\label{soccerball_scale}
\Sigma^\mu\Sigma_\mu=\Pi^\mu\Pi_\mu+\Gamma^\mu\Gamma_\mu+2\Pi^\mu\Gamma_\mu=
2(1+\gamma'_p\gamma'_q)m^2c^2+2\gamma'_p\gamma'_qp^\alpha q_\alpha.
\end{equation}
The last term in Eq. (\ref{soccerball_scale}) is strictly negative, however we notice that for known particles and for the energy scale reachable at present, is extremely small compared to the other terms. It is then evident that the norm of $\Sigma$, which is made of two particles, is bigger then $m^2c^2$. So we can define the scale of the deformations which affect the composite body as
\begin{equation}
\Sigma^\mu\Sigma_\mu=m^{*2}c^2,
\end{equation}
where $\displaystyle m^{*2}=2(1+\gamma'_p\gamma'_q)m^2+2\gamma'_p\gamma'_q\frac{p^\alpha q_\alpha}{c^2}$, which roughly grows faster then the number of constituents. We conclude this section by noting that the same computation made for bodies composed of more then two particles would lead to even bigger norms, which in the case of macroscopic bodies would be proportional at least to the Avogadro number. Thus, according to our interpretation, this solves the soccer ball problem in our framework too.

\section{Conclusions}
In this paper, we have shown a procedure to build up a DSR theory once its fundamental postulates are fixed. Clearly, a debate on the way we have chosen such postulates (in particular the last one) is unavoidable. Moreover, such a formal construction with the emergence of a fifth dimension could be seen as just an academic exercise. Nonetheless, to the best of our knowledge, the proposal of DSR here reported is the first which has altogether a well-grounded geometrical foundation, a right special relativistic limit, and a reasonable physical interpretation. The first of these is a natural requirement as long as we want to generalize a theory with deep geometrical foundations such as special relativity, the second is a necessary condition to satisfy for a theory which hopes to be predictive, and the third is a test for the coherence of the whole framework.

Despite the fact that the emergence of a fifth dimension could seem to be a radical proposal, the justification for its introduction can be found in the fact that both $c$ and $m$ share the same logical role: being observer-independent constants. Let us specify this better: Sec. \ref{intuition1} shows a formal procedure, justified by the accuracy of the result, to derive the special relativity directly from its postulates, enforcing them in the context of Galilean relativity. This procedure gives rise to the intuition of considering time as an extra dimension to add to the three of classical mechanics. We argue that, from a formal point of view, the introduction of a second observer-independent constant besides the speed of light in vacuum c, as required by theories such as DSR, should be treated in the same way. From this perspective, the introduction of one more dimension appears to be the natural consequence of having two observer-independent constants instead of one.

It is worth noting that the approach here proposed and Relative Locality, despite using totally different frameworks, both end up with a relative concept of spacetime. In fact, in the former, a certain spacetime is just a slice of the whole $\mathcal B$ space, taken at a particular value of the $\chi^4$ axis, while in the latter, different spacetimes are seen as different tangent spaces to the momentum space, which thus has assumed a more fundamental role (cf. Refs.\cite{relativelocality1,relativelocality2}).

In concluding this paper, we must, however, underline that, at this stage of development, the corrections introduced by our theory to the predictions of special relativity (see Sec. \ref{the_propagation}) are extremely small [$\mathcal O(m^{-2})$]; thus, a direct test seems to be reachable only far in the future.

\section*{Acknowledgements}
I would like to thank professor Giovanni Montani for his encouragements and useful comments.

\appendix

\section{Geodesic distance over $\mathcal H$}\label{appendix1}
In this appendix, we show explicitly how to compute the geodesical distance from the origin of the hyperboloid $\mathcal H$, defined by the relation in Eq. (\ref{Einstenian_dispersion_relation}), to a generic point upon it. In order to do this, we will use a rescaled version of the $\Pi_\mu$’s and not the $p_\alpha$’s since, as will become clear in a while, this choice will make all computations easier. However, this choice should not cause any concern, since a geodesic distance is a geometric invariant, so the coordinate system used to compute it does not affect the result. We will explicitly state the dependence over $p$ in the end.
\\

Let us rescale, then, the 5-momenta defining $\displaystyle\tilde \Pi_\mu=\frac{\Pi_\mu}{mc}$ so that Eq. (\ref{Einsteinian-Pi-vs-p}) becomes
\begin{equation}\label{Einsteinian-Pi-vs-p-rescaled}
\tilde \Pi_\mu=
\left(\frac{p_\alpha/mc}{\sqrt{1+\frac{p^\alpha p_\alpha}{m^2c^2}}},\frac{1}{\sqrt{1+\frac{p^\alpha p_\alpha}{m^2c^2}}}\right),
\end{equation}
with $\left[\tilde\Pi_\mu\right]=1$; the dispersion relation [Eq. (\ref{Einstenian_dispersion_relation})], which defines $\mathcal H$, becomes
\begin{equation}\label{Einstenian_dispersion_relation-rescaled}
\tilde \Pi^\mu\tilde \Pi_\mu=1;
\end{equation}
and the origin of the rescaled hyperboloid, which we will call $\tilde{\mathcal H}$, becomes $(0,0,0,0,1)$.

In order to compute the geodesic distance over $\tilde{\mathcal H}$, we use the same trick used in Ref. \cite{spinning1}: we exploit the properties of the embedding space $\mathcal B$, which is flat, noting that a geodesic over $\tilde{\mathcal H}$ can be described by the Lagrangian
\begin{equation}\label{Mass-shell_lagrangian}
L=\dd{\tilde\Pi^\mu}{\rho}\dd{\tilde\Pi_\mu}{\rho}+\omega(\tilde \Pi^\mu\tilde \Pi_\mu-1),
\end{equation}
where $\rho$ is the parameter over the geodesic and $\omega$ is a Lagrange multiplier enforcing the constraint on the trajectory of the geodesic $\tilde\Pi_\mu(\rho)$. The Lagrange equations obtained from Eq. (\ref{Mass-shell_lagrangian}) are
\begin{gather}
\ddot{\tilde\Pi}^\mu-\omega\tilde\Pi^\mu=0, \label{Mass-shell_geodesic_eq_of_motion}\\
\tilde \Pi^\mu\tilde \Pi_\mu-1=0 \label{Mass-shell_constraint}
\end{gather}
with $\displaystyle\dot{\tilde\Pi}_\mu=\dd{\tilde\Pi}{\rho}$. In order to solve Eq. (\ref{Mass-shell_geodesic_eq_of_motion}), we have to distinguish the cases $\omega>0$, $\omega<0$ and $\omega=0$. If $\omega>0$, the solution of Eq. (\ref{Mass-shell_geodesic_eq_of_motion}) is
\begin{equation}\label{Mass-shell_willbespacelikegeodesic}
\Pi^\mu(\rho)=A^\mu\cos(\sqrt{\omega}\rho)+B^\mu\sin(\sqrt{\omega}\rho).
\end{equation}
Imposing the constraint in Eq. (\ref{Mass-shell_constraint}), we obtain that
\begin{equation}
A^\mu A_\mu\cos^2(\sqrt{\omega}\rho)+B^\mu B_\mu\sin^2(\sqrt{\omega}\rho)+ 2A^\mu B_\mu\sin(\sqrt{\omega}\rho)\cos(\sqrt{\omega}\rho)=1,
\end{equation}
which implies the conditions
\begin{equation}
\begin{array}{ll}
A^\mu A_\mu=B^\mu B_\mu=1,	&	A^\mu B_\mu=0.
\end{array}
\end{equation}

If $\omega<0$, the solution of Eq. (\ref{Mass-shell_geodesic_eq_of_motion}) is
\begin{equation}\label{Mass-shell_willbetimelikegeodesic}
\Pi^\mu(\rho)=A^\mu e^{-\sqrt{|\omega|}\rho}+B^\mu e^{\sqrt{|\omega|}\rho}.
\end{equation}
Imposing the constraint in Eq. (\ref{Mass-shell_constraint}), we obtain the equation
\begin{equation}
A^\mu A_\mu e^{-2\sqrt{|\omega|}\rho}+B^\mu B_\mu e^{2\sqrt{|\omega|}\rho}+ 2A^\mu B_\mu=1,
\end{equation}
which implies the conditions
\begin{equation}
\begin{array}{ll}
A^\mu A_\mu=B^\mu B_\mu=0,		&\displaystyle	A^\mu B_\mu=\frac{1}{2}.
\end{array}
\end{equation}

If $\omega=0$, the solution of Eq. (\ref{Mass-shell_geodesic_eq_of_motion}) is
\begin{equation}\label{Mass-shell_willbelightlikegeodesic}
\Pi^\mu(\rho)=A^\mu\rho+B^\mu.
\end{equation}
Imposing the constraint in Eq. (\ref{Mass-shell_constraint}) we obtain the equation
\begin{equation}
A^\mu A_\mu\rho^2+B^\mu B_\mu+ 2A^\mu B_\mu\rho=1,
\end{equation}
which implies the conditions
\begin{equation}
\begin{array}{ll}
A^\mu A_\mu=A^\mu B_\mu=0,		&	B^\mu B_\mu=1.
\end{array}
\end{equation}

Now we want to understand which geodesics are timelike, spacelike, and lightlike. In fact, we are interested in computing the geodesic distances of timelike and lightlike geodesics, since we expect the mass-shell relation to be a deformation of the special relativistic one. So we need to analyze the sign of the modulus of the tangent vectors to the geodesics we found so far. In order to compute this value, we can use the tangent vectors as seen by the embedding space $\mathcal B$ using the embedding metric $g$. Thus, once the constraint is taken into account, the modulus of the tangent vectors is simply $\dot{\tilde\Pi}^\mu(\rho)\dot{\tilde\Pi}_\mu(\rho)$. If $\omega>0$, we find
\begin{equation}
\dot{\tilde\Pi}^\mu(\rho)\dot{\tilde\Pi}_\mu(\rho)=\omega,
\end{equation}
so the geodesic is spacelike\footnote{We recall that $g=diag(-1,1,1,1,1)$.}. If $\omega<0$, we find
\begin{equation}\label{Mass-shell_tangenttimelikegeodesicmodulus}
\dot{\tilde\Pi}^\mu(\rho)\dot{\tilde\Pi}_\mu(\rho)=-|\omega|,
\end{equation}
so the geodesic is timelike. If $\omega=0$, we find
\begin{equation}\label{Mass-shell_lightlikeparticlescondition}
\dot{\tilde\Pi}^\mu(\rho)\dot{\tilde\Pi}_\mu(\rho)=0,
\end{equation}
so the geodesic is lightlike. In light of these results, we discard the geodesic in Eq. (\ref{Mass-shell_willbespacelikegeodesic}), focusing first on computing the timelike geodesic distance using the relation in Eq. (\ref{Mass-shell_tangenttimelikegeodesicmodulus}). Assume we are analyzing the timelike geodesic going out from the origin of $\tilde{\mathcal H}$ ($p^\alpha=0$) and arriving at a point with coordinates (written in the embedding space) $\displaystyle\bar{\tilde \Pi}^\mu=\left(\frac{\bar p^\alpha/mc}{\sqrt{1+\frac{\bar p^\alpha \bar p_\alpha}{m^2c^2}}},\frac{1}{\sqrt{1+\frac{\bar p^\alpha \bar p_\alpha}{m^2c^2}}}\right)$. Using the definition of length of a curve, it holds that
\begin{equation}\label{Mass-shell_timelikelenght}
D(0,\bar{\tilde \Pi})=\int_0^1d\rho\sqrt{|\dot{\tilde\Pi}^\mu(\rho)\dot{\tilde\Pi}_\mu(\rho)|}=\sqrt{|\omega|},
\end{equation}
where we use a parametrization over the geodesic such that $p^\alpha(0)=0$ and $p^\alpha(1)=\bar p^\alpha$. From Eq. (\ref{Mass-shell_timelikelenght}) and from the observations done at the beginning of this appendix, it follows that
\begin{equation}\label{Mass-shell_distance_equalities}
D(0,\bar{\tilde \Pi})=\frac{D(0,\bar \Pi)}{mc}=\frac{D(0,\bar p)}{mc}=\sqrt{|\omega|}.
\end{equation}
Using now Eq. (\ref{Mass-shell_willbetimelikegeodesic}) together with the constraint in Eq. (\ref{Mass-shell_constraint}),  we obtain that $\tilde\Pi^\mu(1)\tilde\Pi_\mu(0)=\cosh\sqrt{|\omega|}$. But we observe that $\tilde\Pi^\mu(0)=(0,0,0,0,1)$ while $\displaystyle\tilde\Pi^\mu(1)=\left(\frac{\bar p^\alpha/mc}{\sqrt{1+\frac{\bar p^\alpha \bar p_\alpha}{m^2c^2}}},\frac{1}{\sqrt{1+\frac{\bar p^\alpha \bar p_\alpha}{m^2c^2}}}\right)$, so
\begin{equation}
\frac{1}{\sqrt{1+\frac{\bar p^\alpha \bar p_\alpha}{m^2c^2}}}=\tilde\Pi^\mu(1)\tilde\Pi_\mu(0)=\cosh\sqrt{|\omega|}=\cosh\left(\frac{D(0,\bar p)}{mc}\right),
\end{equation}
where in the last equality we used Eq. (\ref{Mass-shell_distance_equalities}). Solving with respect to $D(0,\bar p)$, we find that
\begin{equation}\label{Mass-shell_geodesictimelikedistance}
D(0,\bar p)=mc\arcosh\left(\frac{1}{\sqrt{1+\frac{\bar p^\alpha \bar p_\alpha}{m^2c^2}}}\right).
\end{equation}
Following the same path for lightlike geodesics we find that
\begin{equation}\label{Mass-shell_geodesiclightlikedistance}
D(0,\bar p)=0,
\end{equation}
and that
\begin{equation}\label{Mass-shell_lightlike_dispersion_relation}
\bar p^\alpha \bar p_\alpha=0.
\end{equation}

In light of the definition in Eq. (\ref{Mass-shell_mass_definition}) and Eqs. (\ref{Mass-shell_geodesictimelikedistance}), (\ref{Mass-shell_geodesiclightlikedistance}) and (\ref{Mass-shell_lightlike_dispersion_relation}), we can summarize the results of this section, using a generic value of the momentum $p_\alpha$, with the relation
\begin{equation}\label{Mass-shell_on_shell_relation_inverse}
m_pc=mc\arcosh\left(\frac{1}{\sqrt{1+\frac{p^\alpha p_\alpha}{m^2c^2}}}\right),
\end{equation}
or its inverse
\begin{equation}\label{Mass-shell_on_shell_relation}
p^\alpha p_\alpha=-m^2c^2\tanh^2\left(\frac{m_p}{m}\right),
\end{equation}
which holds for both timelike and lightlike geodesics.

\section{Derivation of the relation (\ref{M5process_manipulated})}\label{appendix2}
In this appendix, we show how to derive Eq. (\ref{M5process_manipulated}) from Eq. (\ref{M5process'}). Starting from Eq. (\ref{M5process'}), let us then compute the following integral:
\begin{equation}\label{M5process_appendix}
\int_{-\tau}^\tau u'd\tau_u=\int_{-\tau}^\tau u\ d\tau_u-\int_{-\tau}^\tau\frac{m_q}{m_p}(v'-v)d\tau_u,
\end{equation}
where $\tau>0$ is the proper time of the particle with 4-velocity $u$ and $\tau_u$ is the parameter of integration. Using now the fact that the particles before and after the collision are in a free motion, and that there exists a bijective map between the proper times of the two particles, we notice that $u$, $u'$, $v$ and $v'$ are all constants with respect to the flow of $\tau$. Moreover, choosing without loss of generality that the proper time of both particles is zero at the collision, we have that $u(\tau_u>0)=v(\tau_u>0)=u'(\tau_u<0)=v'(\tau_u<0)=0$. Using these conditions, the result of the integral [Eq. (\ref{M5process_appendix})] is
\begin{equation}
u'\tau=u\tau-\frac{m_q}{m_p}(v'-v)\tau.
\end{equation}
Finally, choosing the spacetime point of the collision to be the origin of our reference frame, we can interpret the left-hand side and the first term of the right-hand side to be the positions of the particles with 4-velocity $u'$ and $u$, respectively; thus
\begin{equation}
x'=x-\frac{m_q}{m_p}(v'-v)\tau,
\end{equation}
where we have called $x'=u'\tau$ and $x=u\tau$. It is worth noting that, since $u(\tau_u>0)=0$, $x$ is the position the particle with 4-velocity $u$ would have if the collision had not occurred.

\end{document}